\documentclass[12pt]{iopart}

\makeatletter
\newenvironment{figurehere}
           {\def\@captype{figure}}
           {}
           \makeatother
           \usepackage{graphicx}
           \usepackage{dcolumn}
           \usepackage{bm}
           \usepackage{epsfig}
           \usepackage{color}
\usepackage{color}
\usepackage{amssymb}

\begin{document}

\title{Phase transitions and entanglement properties in spin-1 Heisenberg clusters with single ion anisotropy}
\author{V. S. Abgaryan$ ^1$, N. S. Ananikian$ ^{1}$, L. N. Ananikyan$ ^1$ and A. N. Kocharian$^{1,2}$}

\address{$^1$ Alikhanyan National Science Laboratory, Alikhanian Br. 2, 0036 Yerevan, Armenia}
\address{$^2$ Department of Physics, California State University, Los Angeles,CA 90032, USA }
\ead{ananik@yerphi.am}
\begin{abstract}
The incipient quantum phase transitions of relevance to non-zero
fluctuations and entanglement are studied in Heisenberg clusters
by exploiting the negativity as a measure  in bipartite and
frustrated spin-1 anisotropic Heisenberg clusters with
bilinear-biquadratic exchange, single-ion anisotropy and magnetic
field. Using exact diagonalization technique it is shown that
quantum critical points signaled by qualitative changes in
behavior of magnetization and particle number, is ultimately
related to microscopic entanglement and collective excitations.
The plateaus and peaks in spin and particle susceptibilities,
define the conditions for a high/low density quantum entanglement
and various ordered phases with different spin (particle)
concentrations.
         \end{abstract}
         \maketitle
% ----------------------------------------------------------------
\section{Introduction}
Entanglement properties for few spins or electrons can display the
general features of large thermodynamic systems and different
measures of entanglement have been defined to understand
QPTs~\cite{Amico,Sach}. A finite spin system is important in the
context of molecular magnetism and spin pairing. A new line of
research points to a connection between the local entanglement in
one-dimensional correlated systems and the existence of a QPTs and
scaling~\cite{Larsson,Yang} relevant to the quantum critical
points (QCPs). Furthermore, such a connection can be exploited to
unveil a fundamental connection between the QCPs in finite-size
small, large clusters~\cite{Nagaoka,PRB} and macroscopic
systems~\cite{Sach}. A particle and spin density fluctuations,
extending the essential properties of entanglement beyond the
conventional framework, have been introduced with the explicit
reference to the phase transitions in canonical and grand
canonical ensembles signaled by a critical behavior in terms of
the energy gaps and susceptibilities~\cite{akocha,physlett}.  The
quantum gas of clusters at the equilibrium gives unprecedented
opportunity to explore exactly these ideas for a quantum dynamics
of spin fluctuations~\cite{Nature1}. While the basic features of
entanglement in spin-${1\over 2}$ systems are by now fairly well
understood~\cite{Wang_Fu}, entanglement properties of larger spin
fermions (or bosons) are less known due to the lack of good
operational measures for high spin entanglement~\cite{Mintert}. A
general classical spin-1 Blume-Emery-Griffiths (BEG)
model~\cite{Grif} has proven to be a useful for description of
liquid-gas, liquid-crystal phase transitions, tricritical and
$\lambda$ points, spontaneous phase
separation~\cite{Wong,Bellini,Wu,Avak}. The integer spin
Heisenberg model exhibits a characteristic spin gap and very rich
phase diagrams~\cite{Babujan,Haldane}.
%\\ However, the closed form
%solution, existing in the Bethe-ansatz only for specific set of
%parameters is difficult to analyze especially at finite
%temperatures without having to resort to various approximations.
%On the contrary,
Exact calculations of thermodynamic and entanglement properties in
finite-size clusters can give an appealing alternative to get
insight into the general features of bipartite and frustrated
systems~\cite{Arnesen1}. Some analytical and numerical studies of
entanglement and negativity in bilinear-biquadratic spin-1
Heisenberg model on dimerized bipartite and frustrated systems
have been performed in \cite{Schollwock,Sun,Guo,Li}. The
entanglement with bilinear-biquadratic Hamiltonian has been
considered for the case of two spins (qubit) in the absence of
crystal field~\cite{Zhou}. One of interesting problems concerning
entanglement is to study the effect of uniaxial single-ion
anisotropy and magnetic fields on negativity. An exact calculation
of entanglement versus longitudinal crystal field and biquadratic
coupling for analyzing the variation of negativity versus
parameters of the spin-1 system have not been attempted either for
ferromagnetic or antiferromagnetic Heisenberg model even for small
bipartite and frustrated clusters. The aim of this work is to
discuss, in a general framework, how microscopic entanglement in
the two- and three-qubit context can be related to a QCPs
characterized by plateaus in peak behavior of the spin revealed in
saddle point singularities on model parameters. We provide a
reinterpretation of the spin and particle susceptibilities near
quantum critical points in terms of the quantum entanglement in a
physically transparent way. Here we adopt negativity to measure the ground state
entanglement for spin-1 systems, to reveal QPTs in terms of negativity. We have two main goals in this paper: The
first is to provide a global view of the most general spin-1
Heisenberg model which  have not been highlighted so far in
minimal clusters. The second is to show that quantum entanglement
exhibits the existence of characteristic plateaus in negativity
related to QPTs. In this paper we perform exact calculations of
entanglement and response functions in spin-1 Heisenberg model
with bilinear-biquadratic exchange interactions in longitudinal
crystal and magnetic fields. The basic principles for calculation
of negativity are introduced in Sect.~\ref{basic}. The ground
state magnetic and entanglement properties in spin-1 Heisenberg
model for ferromagnetic and antiferromagnetic couplings are given
in Sect.~\ref{isotropic}. The negativity analyzes in absence and
presence of magnetic field are given in Sect.~\ref{absence} and
Sect.~\ref{magnetic} respectively. The effect of nonlinear
interaction is studied in Sect.~\ref{quadruple}). Results for
frustrated trimer are presented in Sect.~\ref{NonBipartite}. The
conclusion is given in Sect.~\ref{conclusion}.
\section{Spin-1 Heisenberg model}~\label{Hamiltonian}
We consider the spin-1 isotropic Heisenberg model in the presence
of magnetic field $B<0$
\begin{eqnarray}
   H & = & \sum _{i=1}^N [J (\vec{S}_{i}\vec{S}_{i+1}) +K (\vec{S}_{i}\vec{S}_{i+1})^2]  + \\
\nonumber
 &&D \sum
_{i=1}^N(S_i^{z})^2+B\sum _{i=1}^N S_i^z.  \label{mu}
\end{eqnarray}
The linear $J$ and nonlinear $K$ terms are the exchange and
quadrupolar interactions. Here we implemented the longitudinal
crystal field $D$, which describes an uniaxial single-ion
anisotropy. In what follows, the crystal field significantly
changes the results on the entanglement. Notice, effective spin
Hamiltonian (\ref{mu}) can be derived from Bose-Hubbard model in
the strong coupling limit. The local spin vector $\vec{S_{i}}$ for
each site has components of the Spin-1 operators
\begin{eqnarray}
&S_{x}&=\frac{1}{\sqrt{2}} \left(\nonumber
\begin{array}{lll}
 0 & 1 & 0 \\
 1 & 0 & 1 \\
 0 & 1 & 0
\end{array}
\right), S_{y}=\frac{1}{\sqrt{2}} \left(
\begin{array}{lll}
 0 & -i & 0 \\
 i & 0 & -i \\
 0 & i & 0
\end{array}
\right),
\\
&S_{z}&= \left(
\begin{array}{lll}
 1 & 0 &  0 \\
 0 & 0 &  0 \\
 0 & 0 & -1
\end{array}
\right).
\end{eqnarray}
Unless otherwise specified, we will consider periodic boundary
conditions, such that
%with imposed periodic boundary condition,
$\vec{S}_{N+1}=\vec{S}_{1}$, where $N$ is the total number of
lattice sites. The sum over lattice sites for crystal field term
with $(S_i^{z})^2$ in (\ref{mu}) can be reduced to the spin
concentration (particle number),
\begin{eqnarray}
 \nonumber
\sum _{i=1}^N(S_i^{z})^2=P-P_0,\label{D}
\end{eqnarray}
where $P_0$ the number of lattice sites with $S_i^{z}=0$. Notice,
the axial anisotropy in many respects is analogous to the chemical
potential $D=-\mu$.
\section{Definitions and Basic}~\label{basic}
At thermal equilibrium, the state of the system is determined by
the density matrix
\begin{equation}
\hat{\rho}(T)={e^{-{H\over k_BT}}\over Z}=\sum_i{e^{-{E_i\over
k_BT}}\over Z}|\psi_i\rangle \langle \psi_i|,
 \label{rho}
\end{equation}
where $E_i$ are the eigenvalue of the $i$-th quantum many body
eigenstate and the partition function is $Z=\sum_i e^{-\beta E_i}$
with $\beta = {1/k_BT}$ ($k_B = 1$). The many-body entanglement is
described by the density operator in
\cite{Arnesen1,Wang,Schliemann,Dur W}. For spin-1 system the
degree of pairwise entanglement, measured in terms of the
negativity $Ne$, can be employed to evaluate the thermal state of
concern~\cite{Vidal}. The negativity of a state $\rho$ is defined
as
\begin{equation}
\textit{Ne}=\sum_{i}\ |\mu_{i}| , \label{negat1}
\end{equation}
where $\mu_{i}$ $\acute{}$s are negative eigenvalues of
$\rho^{T_{1}}$ and $T_1$ denotes the the partial pairwise
transpose with respect to the first system, i.e., for bipartite
system in state $\rho$ it is defined as
\begin{equation}
\langle i_{1},j_{2}|\rho^{T_{1}}|k_{1},l_{2}\rangle \equiv\langle
k_{1},j_{2}|\rho|i_{1},l_{2}\rangle,
\end{equation}
for any orthonormal but fixed basis. Definition (\ref{negat1}) is
equivalent to
\begin{equation}
\textit{Ne}=\frac{||\rho^{T_{1}}||_{1}-1}{2},
\end{equation}
where $||\rho^{T_{1}}||_{1}$ is trace norm of $\rho$
($\rho=Tr\sqrt{\rho^{\dag}\rho}$). For unentangled states
negativity vanishes, while \textit{Ne}$>$0 gives a computable
measure of thermal entanglement.\\
As thermodynamical characterisation we have used the responses of the thermodynamical
potential with respect to $D$ and $B$ which are follows
\begin{eqnarray}
P=\left\langle (S^{z})^{2}\right\rangle={{\frac {\partial {F}} {\partial D}}}, \
\ \ \ \ \left\langle S^{z}\right\rangle={{\frac {\partial F}
{\partial B}}}.\label{spin}
\end{eqnarray}
Here $F$ is the free energy $F=-T ln Z$ and ${\left\langle
....\right\rangle}$ indicates averaging performed within a
canonical ensemble. The responses for the first derivatives of the
thermodynamic potential with respect to $D$ and $B$ provide exact
expressions for particle $\chi_D$ and spin $\chi_B$
susceptibilities:
\begin{eqnarray}
{\chi_D}={{\frac {\partial {P}}
{\partial D}}}, \ \ \ \ \ \ \ \ \chi_B={{\frac {\partial
\left\langle S^{z}\right\rangle} {\partial B}}}\label{spin1}
\end{eqnarray}
\section{Results}~\label{Res}
\subsection{Entanglement and magnetic
properties of spin-1 isotropic Heisenberg dimer}~\label{isotropic}
In this section, we consider Hamiltonian in case of $N=2$, namely
%XXX
Heisenberg model. In the two-qubit case, we diagonalize the
Hamiltonian, and obtain the eigenvalues
\begin{eqnarray}
    \nonumber E_{1}&=&-2(B-J-K-D),\quad E_{2}=-2(J-K-D),\\\nonumber
    E_{3}&=&2(B+J+K+D),\quad E_{4}=-B-2J+2K+D, \\\nonumber
    E_{5}&=&B-2J+2K+D,\quad E_{6}=-B+2J+2K+D,\\\nonumber
    E_{7}&=&B+2J+2K+D,\quad E_{8}=-J+5K+D-\lambda_{0} \\
    E_{9}&=&-J+5K+D+\lambda_{0},\\\nonumber
 \label{eigenvals}
\end{eqnarray}
and corresponding eigenvectors
\begin{eqnarray}
    \nonumber &|\psi_{1}\rangle&=|-1, -1\rangle,\quad |\psi_{2}\rangle=\frac{1}{\sqrt{2}}(|-1,1\rangle-|1,-1\rangle), \\\nonumber
    &|\psi_{3}\rangle&=|1, 1\rangle,\quad |\psi_{4}\rangle=\frac{1}{\sqrt{2}}(|-1,0\rangle-|0,-1\rangle), \\\nonumber
    &|\psi_{5}\rangle&=\frac{1}{\sqrt{2}}(|0,1\rangle-|1,0\rangle),|\psi_{6}\rangle=\frac{1}{\sqrt{2}}(|-1,0\rangle+|0,-1\rangle),\\\nonumber
    &|\psi_{7}\rangle&=\frac{1}{\sqrt{2}}(|0,1\rangle+|1,0\rangle),\\\nonumber
    &|\psi_{8}\rangle&=\frac{1}{\sqrt{2+\lambda_{1}^2}}(|1,-1\rangle+\lambda_{1}|0,0\rangle+|-1,1\rangle),\\
    &|\psi_{9}\rangle&\frac{1}{\sqrt{2+\lambda_{2}^2}}(|1,-1\rangle+\lambda_{2}|0,0\rangle+|-1,1\rangle)
    \label{eigenstates}
\end{eqnarray}
where,
$\lambda_{0}=\sqrt{9(J-K)^2-2(J-K)D+D^2},\lambda_{1}=\frac{J-K-D-\lambda_{0}}{2(J-K)},\lambda_{2}=\frac{J-K-D+\lambda_{0}}{2(J-K)},$
and $|i,j\rangle$ ($i=-1,0,1$ and $j=-1,0,1$) are the eigenvectors
of $S_i^{z}S_{i+1}^{z}$. According to Schmidt theorem
$|\psi_5\rangle$ and $|\psi_7\rangle$ are not entangled and the
maximum entangled states can be only $|\psi_8\rangle$ or
$|\psi_9\rangle$.
The partial transpose density matrix of the thermal state
$\rho(T)$ at equilibrium is
\begin{equation}
\nonumber \rho^{T_{1}}=\frac{1}{Z}
    \left(
\begin{array}{lllllllll}
 \omega ^- & 0 & 0 & 0 & \chi ^- & 0 & 0 & 0 & \Xi ^- \\
 0 & \chi ^+ & 0 & 0 & 0 & \Omega  & 0 & 0 & 0 \\
 0 & 0 & \Xi ^+ & 0 & 0 & 0 & 0 & 0 & 0 \\
 0 & 0 & 0 & \chi ^+ & 0 & 0 & 0 & \Omega  & 0 \\
 \chi ^- & 0 & 0 & 0 & \Lambda  & 0 & 0 & 0 & \zeta ^- \\
 0 & \Omega  & 0 & 0 & 0 & \zeta ^+ & 0 & 0 & 0 \\
 0 & 0 & 0 & 0 & 0 & 0 & \Xi ^+ & 0 & 0 \\
 0 & 0 & 0 & \Omega  & 0 & 0 & 0 & \zeta ^+ & 0 \\
 \Xi ^- & 0 & 0 & 0 & \zeta ^- & 0 & 0 & 0 & \omega ^+
\end{array}
\right),
\end{equation}
where
\begin{eqnarray}
    \nonumber  &\omega^{\pm}&=e^{\frac{2(\pm B-D-J-K)}{T}},\quad \chi^{\pm}=\frac{1}{2} e^{-\frac{B+2 (J+K )+D }{T}} \left(1 \pm e^{\frac{4J}{T}}\right),\\
    \nonumber  &\zeta^{\pm}&=\frac{1}{2} e^{\frac{B-2 (J+K )-D }{T}} \left(1 \pm
    e^{\frac{4J}{T}}\right),\\
    \nonumber  &\Xi^{\pm}&=\pm \frac{1}{2}
    e^{\frac{2(J-D-K)}{T}}+\\\nonumber
    &&\frac{e^\frac{J-5K-D}{T}(\lambda_{0}\cosh\frac{\lambda_{0}}{T}+(J-K-D)\sinh\frac{\lambda_{0}}{T}}{2\lambda_{0}},\\
    \nonumber
    &\Omega&=\frac{2e^\frac{J-5K-D}{T}(K-J)\sinh\frac{\lambda_{0}}{T}}{\lambda_{0}},\\
   \nonumber &\Lambda&=\frac{e^\frac{J-5K-D}{T}(\lambda_{0}\cosh\frac{\lambda_{0}}
   {T}-(J-K-D)\sinh\frac{\lambda_{0}}{T}}{\lambda_{0}},
\end{eqnarray}
here the partition function is
\begin{eqnarray}
 \nonumber
    Z&=&e^{-\frac{2 (D+J)+5 K}{T}}(2 e^{\frac{D+3 K}{T}} (1+e^{\frac{4 J}{T}}) \cosh(\frac{B}{T})+\\
 \nonumber   &&e^{\frac{4 J+3 K}{T}}+2 e^{\frac{3 K}{T}}\cosh(\frac{2 B}{T})+2 e^{\frac{D+3 J}{T}} \cosh (\frac{\lambda
 _0}{T})).
\end{eqnarray}
%We do not present the result for negativity here due to its
%complexity.
\subsection{Spin-1 in zero magnetic field}~\label{absence}
Here we analyze the effects of crystal field on the ground state
entanglement in the spin-1 Heisenberg model (\ref{mu}) at zero
field ($B=0$). The (quadrupole) particle number and negativity
plots in figures~\ref{fig particle_D} {a} and {b} are both
asymmetric as function of $D$ for ferro $J>0$ and
antiferromagnetic $J<0$ couplings. The monotonic behavior of $P$
versus $D$ in figure~\ref{fig particle_D} {a} signals a smooth
character of the phase transition. Note that this smooth bosonic
behavior of spin concentration $P$ versus $D$ is contrasted from
the (step-like) abrupt fermionic change in the electron number as
a function of the chemical potential in \cite{PRB}. At
infinitesimal $T \rightarrow 0$ the variation of negativity
figure~\ref{fig particle_D}b versus $D$ for antiferromagnetic case
($J>0$) is non monotonic. So, for $D=0$, we have a highest
possible entanglement and $\psi_{8}$ is a ground state of the
system. The system for $J<0$ displays two distinct phases: one
separable and other entangled. For positive $D$ region the
negativity for $J>0$ is more than for $J<0$. For $D=0$ the ground
state is five fold degenerate (i.e., it is a mixture of
$\psi_{1},\psi_{3},\psi_{6},\psi_{7},\psi_{8}$ states) with zero
negativity. At infinitesimal $D\to +0$ the system is entangled in
the pure state, $\psi_{8}$. For $D<0$ the ground state at $J<0$ is
double degenerate with a mixture of $\psi_{1}$ and $\psi_{3}$
states.
\begin{figurehere}
\begin{center}
\begin{tabular}{cc}
{\small (a)}&{\small (b)}\\
\includegraphics[width=35mm]{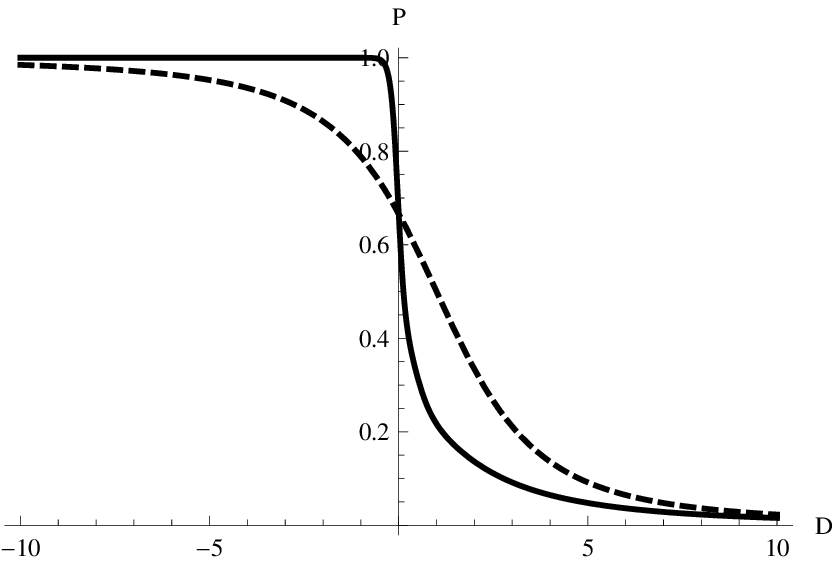}&\includegraphics[width=35mm]{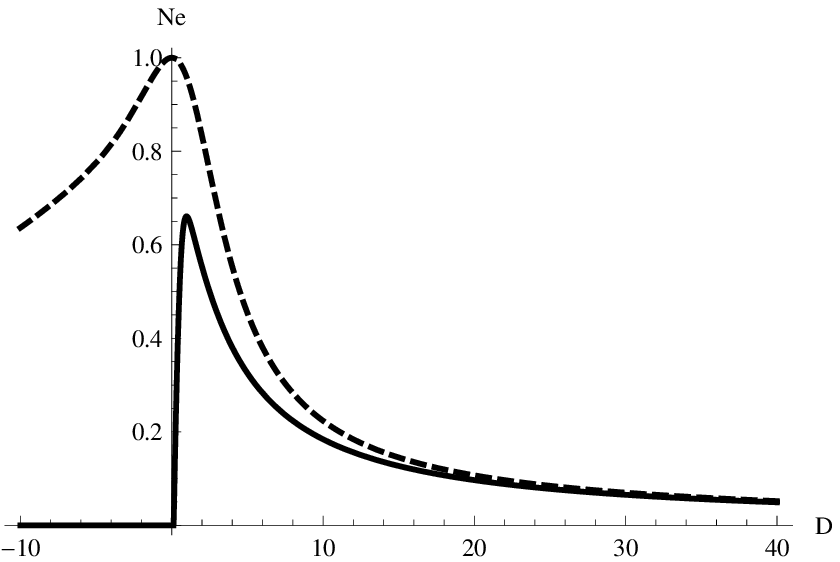}\\
\end{tabular}
\caption {\small {The density variation of a) the particle number
$P$ and b) the negativity $Ne$ versus $D$ for antiferromagnetic,
$J=1$ (dashed) and ferromagnetic, $J=-1$ (solid) cases}}
 \label{fig particle_D}
 \end{center}
\end{figurehere}
Notice, these states are separable (can be factorized), and
therefore, according to definition, these quantum states are
without entanglement. Thus, the entanglement in $D<0$ region can
be used to detect quantum correlations in antiferromagnetic case,
which are absent for ``classical" ferromagnetic.
 \begin{figurehere}
           \begin{center}
             \psfig{figure=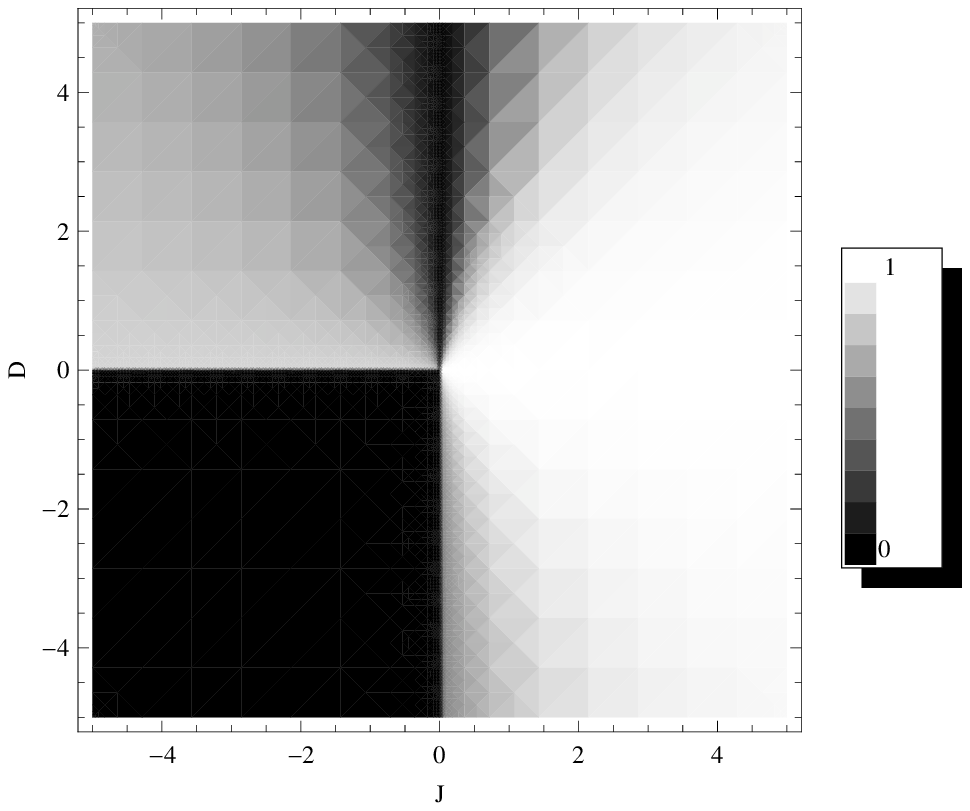,width=45mm}
           \end{center}
           \caption{\small {The density of negativity $Ne$ versus of $J$ and $D$. The crystal
field $D$ enhances the entanglement at
$J<0$.}}\label{hbound02}\vspace{0.01cm}
         \end{figurehere}
The negativity is non monotonous function with one maxima at $D=0$
for $J>0$ and in a close vicinity to origin at $J<0$. The magnetic
and quadrupole susceptibilities, i.e $\chi_B$ $\chi_D$, allows to
distinguish the ordered and disordered phases in the case of
broken-symmetry at QPTs. Figure~\ref{hbound02} shows the pure
(extremal) and mixed (non-extremal) quantum states. Disentangled
dark region in ferromagnetic case corresponds to the plateau-like
behavior in zero (spin) magnetic susceptibility $\chi_0={{\frac
{\partial \left\langle s^{z}\right\rangle} {\partial h}}}|_{B\to
0}$ versus $J$ and $D$ plane in figure~\ref{chi} {a} at $J<0$.
\begin{figurehere}
\begin{center}
\begin{tabular}{cc}
{\small (a)}&{\small (b)}\\
\includegraphics[width=35mm]{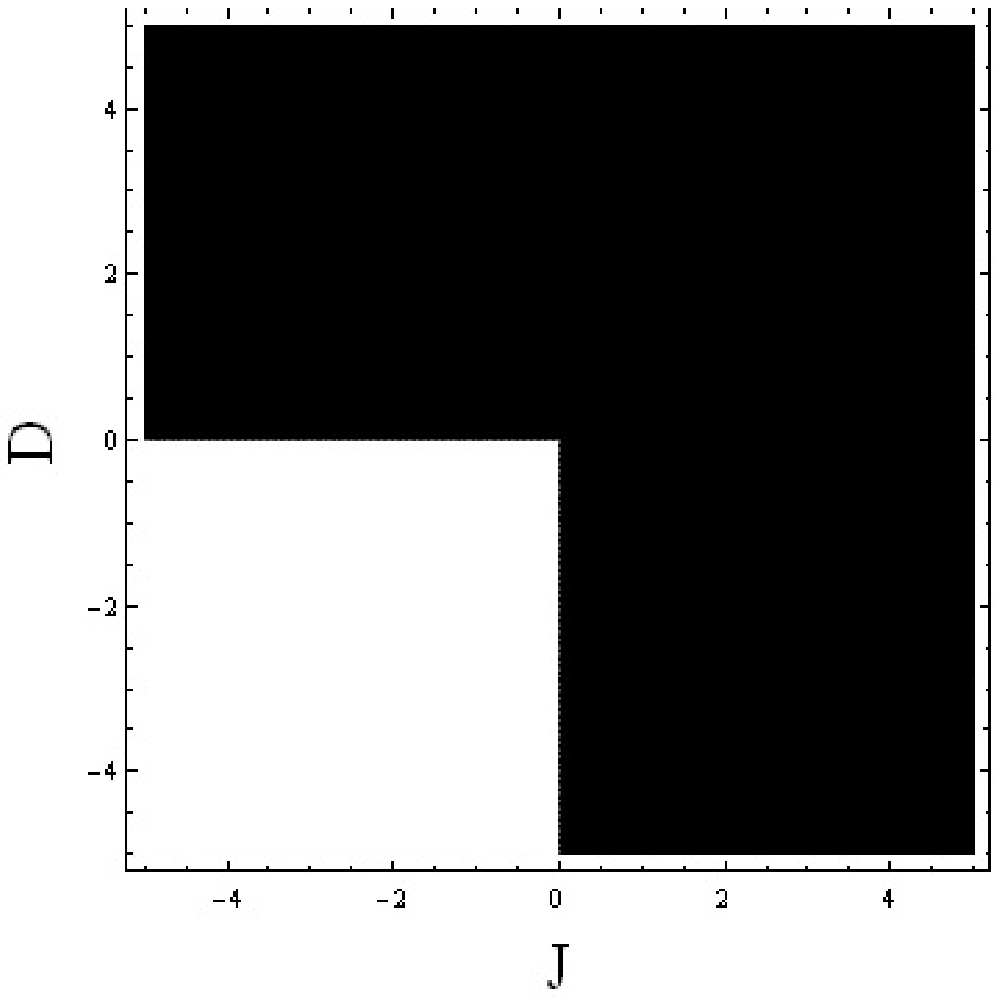}&\includegraphics[width=35mm]{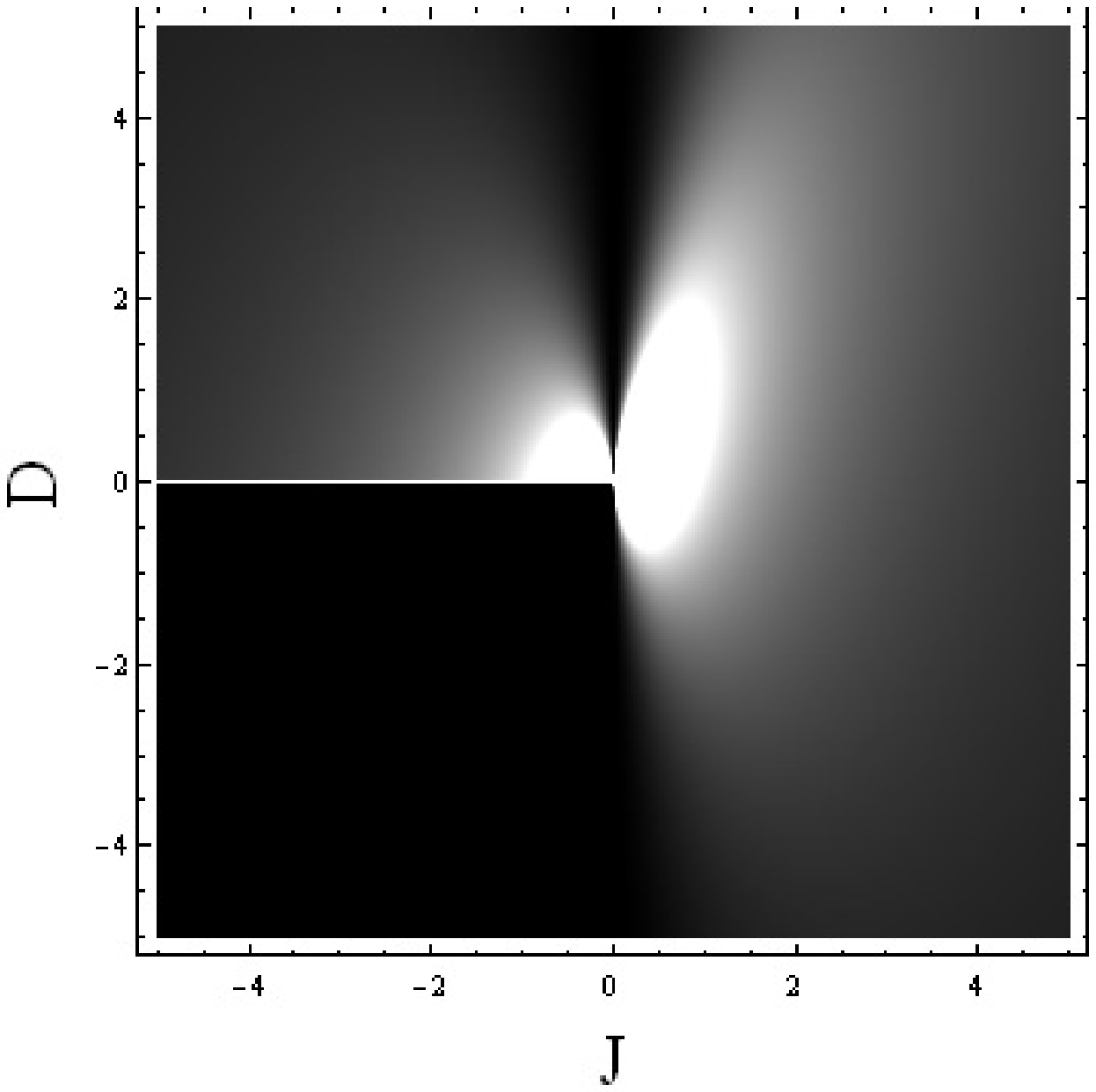}\\
\end{tabular}
\caption {The densities for a)  zero field (magnetic) spin
susceptibility and b) particle susceptibility versus $J$ and $D$.}
 \label{chi}
 \end{center}
\end{figurehere}
 The high density magnetic (spin) susceptibility in white sector corresponds to the
low density of negativity in figure~\ref{hbound02}. The strong
enhancement of negativity along the line $D=0$ is relevant to the
observed peaks in the particle susceptibility, ${\chi_D}={{\frac
{\partial {\left\langle {(S^z)^2}\right\rangle}} {\partial D}}}$
in figure~\ref{chi} {a}. The various regions seen for (density)
negativity in figure~\ref{hbound02} are reproduced in density of
quadrupole susceptibility in figure~\ref{chi} {b} versus $D$ and
$J$. Similarly, the phase diagram in $K-J$ space in the absence of
$D$ and $B$ fields shows the degree of entanglement and phases due
to effect of nonlinearity on the eigenvalues and eigenvectors in
(\ref{eigenvals}). For example, the $J=K$ line separates the
maximum entangled and non-entangled phases for ferromagnetic
coupling, while the $J=3K$ line is a boarder between entangled and
new less-entangled phases for antiferromagnetic coupling.
\begin{figurehere}
\begin{center}
\begin{tabular}{cc}
{\small (a)}&{\small (b)}\\
\includegraphics[width=40mm]{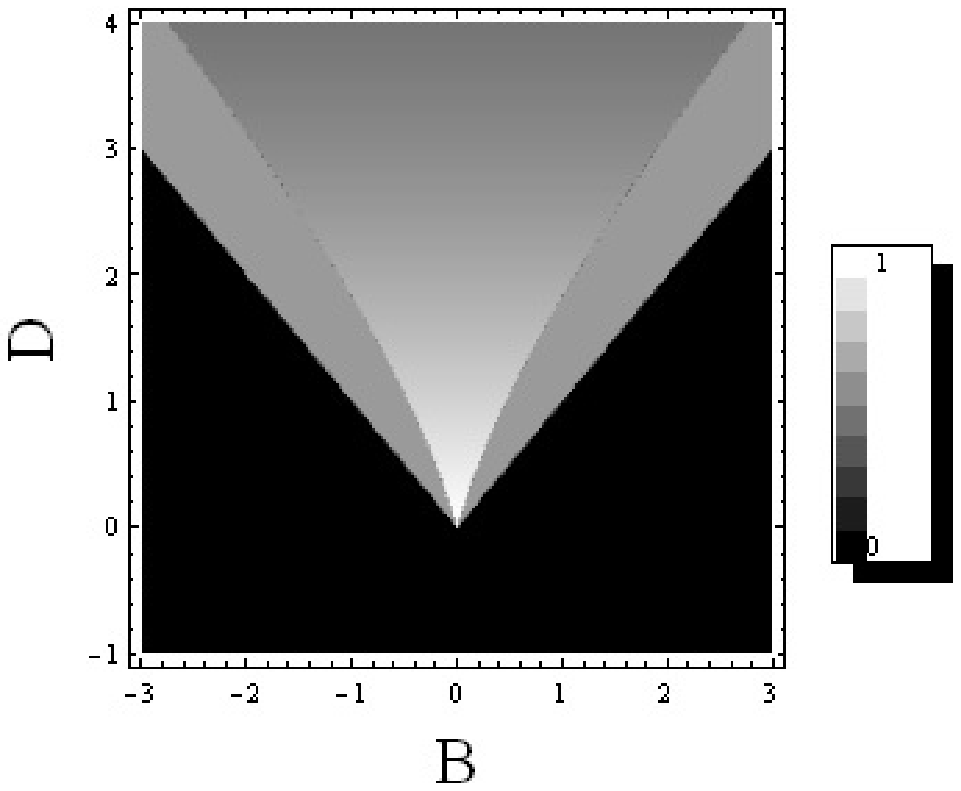}&\includegraphics[width=32mm]{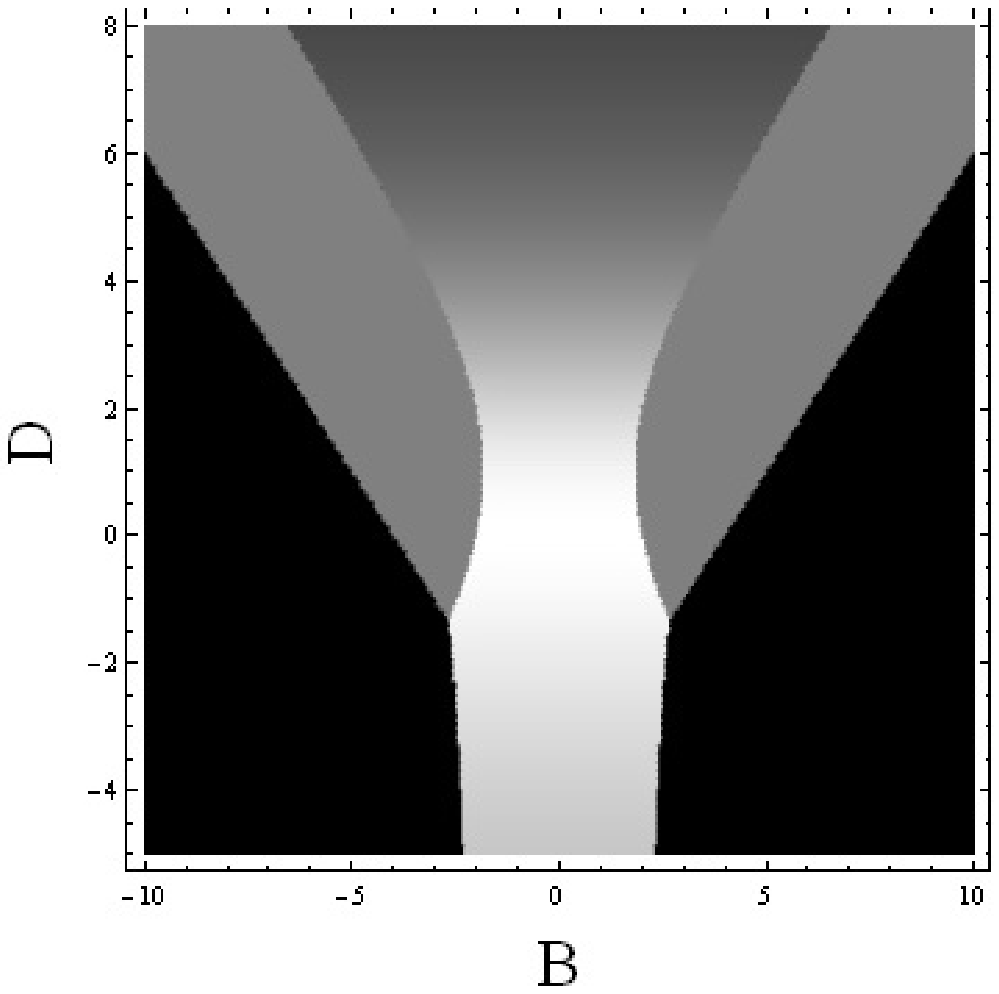}\\
\end{tabular}
\caption {\small {The density plot for negativity dependencies on
crystal $D$ and magnetic $B$ fields at $T=0$ for (a) $J=-1$ and
(b) $J=1$. For both (antiferromagnetic and ferromagnetic) cases
there are more than three phases, which indicates the possible
existence of triple or tricritical points.}}
 \label{hbound03}
 \end{center}
\end{figurehere}
We find that for $K>0$, the line $J=0$ as before separates non
entangled and maximum entangled phases. The maximum entanglement,
which exists for $J<0$ at $K<J$ and for $J>0$ at $J>3K$,
corresponds to observed condition for bose condensation of
unpolarized $Na$ atoms on optical lattice~\cite{Stenger}.
\subsection{Effects of magnetic field}~\label{magnetic}
Magnetic field $B$ partially removes the ground state degeneracy
%decrease degenerations
and in figure~\ref{hbound03} one can see the presence of new phase
boundaries. The entanglement properties of the excited states are
independent from those in the ground states. Also we found that
the pairwise entanglement decreases from ground state to excited
states, i.e., the more excited the system, the less the
entanglement. In ferromagnetic case for $D=0$ and $B=0$ point
there is a maximum entangled state. In figures~\ref{hbound03} {a}
for $J=-1$ and ~\ref{hbound03} {b} for $J=1$, the energies are
measured with respect to $|J|$, which is set to 1. When $D<|B|$
the system is in $\psi_{1}$ or $\psi_{3}$ state, which is
non-entangled. For fixed $B$ the two consecutive phase transitions
take place by increasing $D$ at $D=|B|$ and $D=\sqrt{1+6 |B|+B^2}
-1$, into $\psi_{6}$ and $\psi_{8}$ ground states correspondingly.
For antiferromagnetic case, the phase diagram is more complex. The
negativity contains the triple point at $|B|=\frac{8}{3}$ and
$D=-\frac{4}{3}$, which implies the presence of various phases,
possible coexistence or phase separation in spin-1 system. When
$D<-\frac{4}{3}$, the line $|B|=-\frac{2}{-2+|B|}-1$ separates
$\psi_{8}$ and $\psi_{1,4}$, i.e. maximum entangled phase from
non-entangled one. For $D>-\frac{4}{3}$ there are three phases in
the ground state: non-entangled state at $D<|B|-4$; the maximum
entangled between $1+\sqrt{B^2+2 |B|-7}$ and $1-\sqrt{B^2+2
|B|-7}$; non saturated entanglement for $\psi_{4,5}$ states. In
figure~\ref{hbound03} there is no any entanglement beyond some
critical field $B_c$ restricting the black region. Also note that
entanglement increases with $D$. Positive $D$ values favor to the
larger entanglement, while $D<0$ shows the tendency toward
non-entangled states with larger total spin. The ground state
diagrams figures~\ref{hbound02}, and~\ref{hbound03} exhibit
quantum critical behavior on the borderlines between various
states with continuously-varying quantum critical points
separating antiferromagnetically ordered distinct phases from from
the non-entangled state (spin liquid phase). These critical lines,
similar to continues QCPs, can be used for the classification of
the many-body ground states of interacting spins and quadrupole
moment in multidimensional parameter space. Dynamic interactions
between the spins strongly renormalize various parameters in the
effective Hamiltonian and, therefore, spin and quarupole momentum
have properties different from a quasiparticle description. As in
QCP~\cite{Sach}, various states along quantum critical boundaries
here are necessarily separated by second order phase transitions
with various entanglement and susceptibility. The quantum critical
(lines) boundaries appear to be a useful for understanding the
formation of various thermodynamic phases in the ground state.
These continues boundaries in thermodynamic phase diagram at
infinitesimal temperature coincide with the corresponding QCPs,
derived from the peaks of magnetic susceptibilities in agreement
with our preliminary analysis (see also \cite{physlett,akocha}).
The boundaries between the various phases are useful for
understanding also the behavior of thermal negativity for
departure to non-zero temperatures.
\begin{figurehere}
\begin{center}
\begin{tabular}{cc}
{\small (a)}&{\small (b)}\\
\includegraphics[width=28mm]{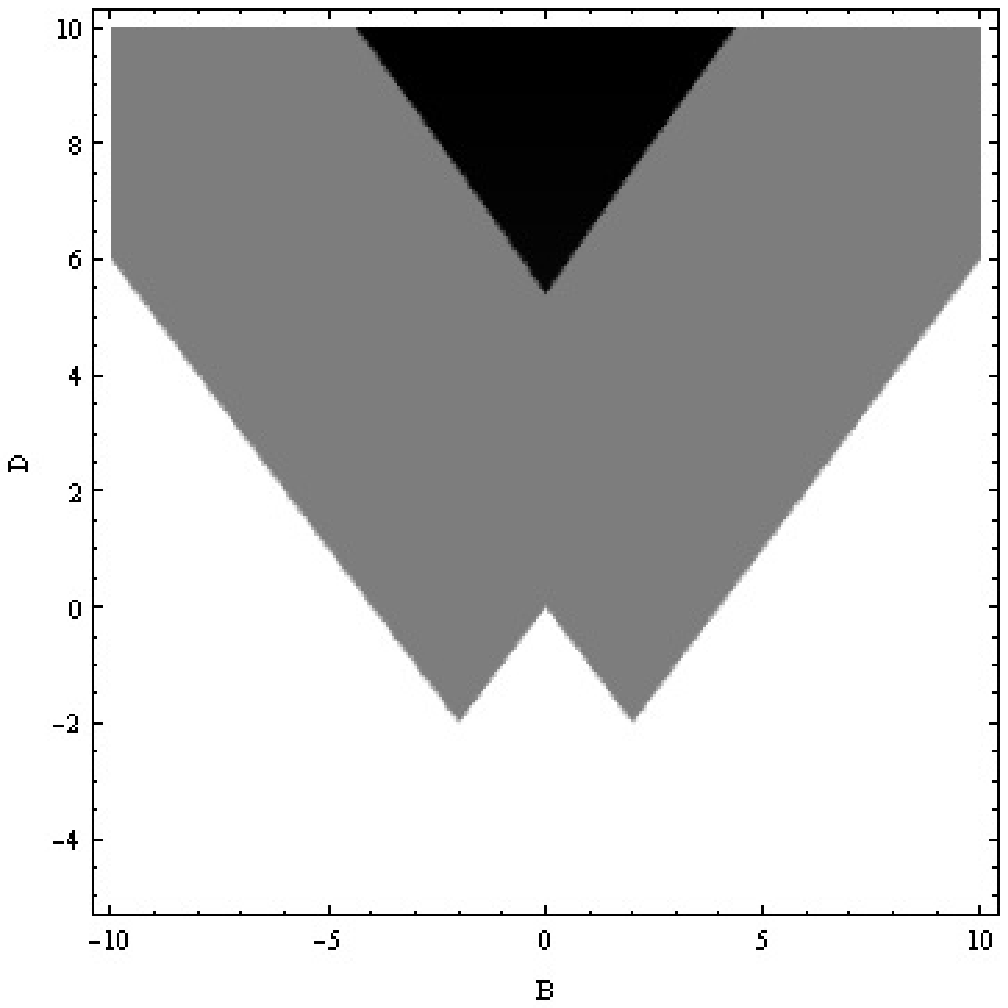}&\includegraphics[width=30mm]{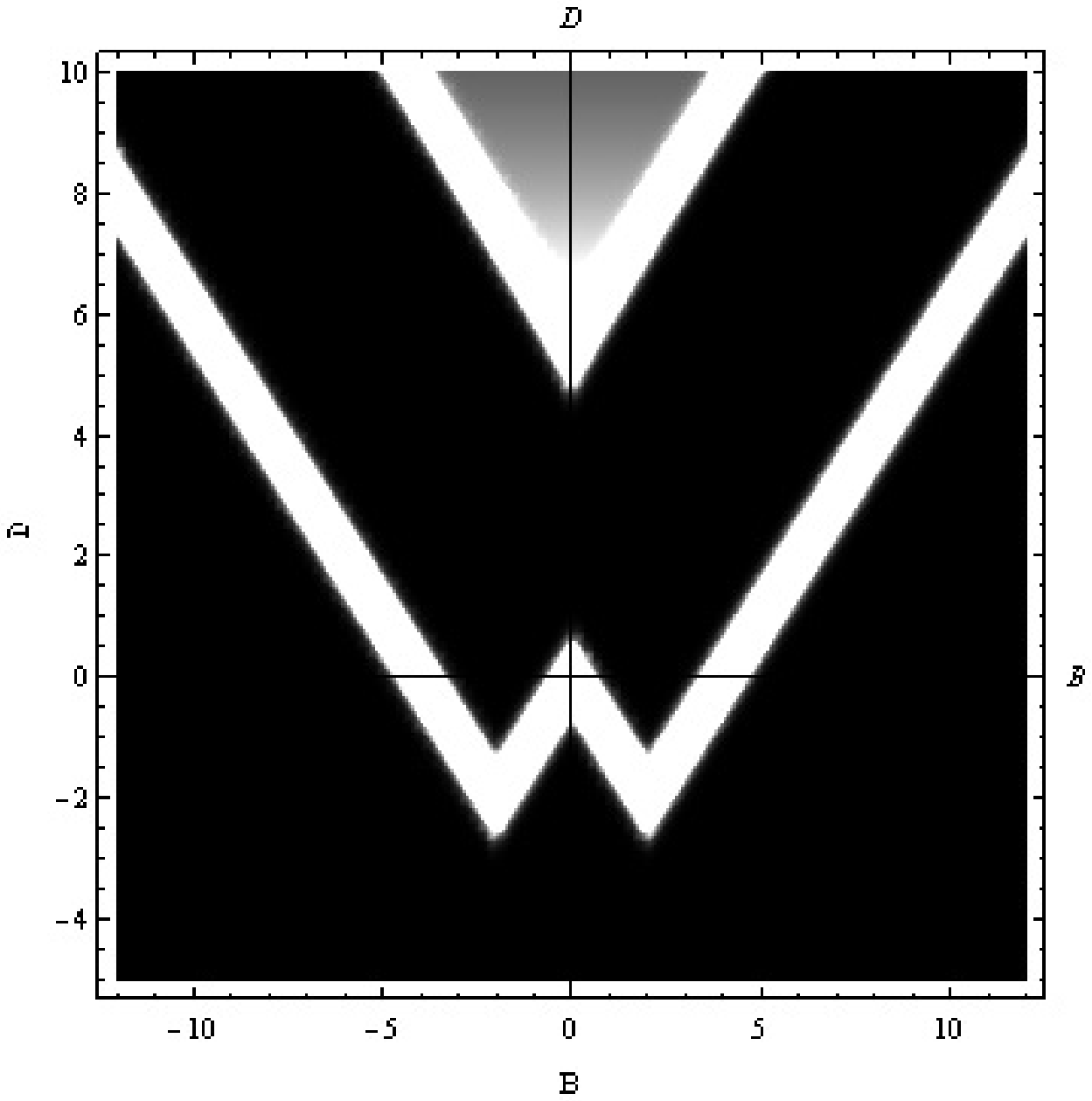}\\
{\small (c)}&{\small (d)}\\
  \includegraphics[width=30mm]{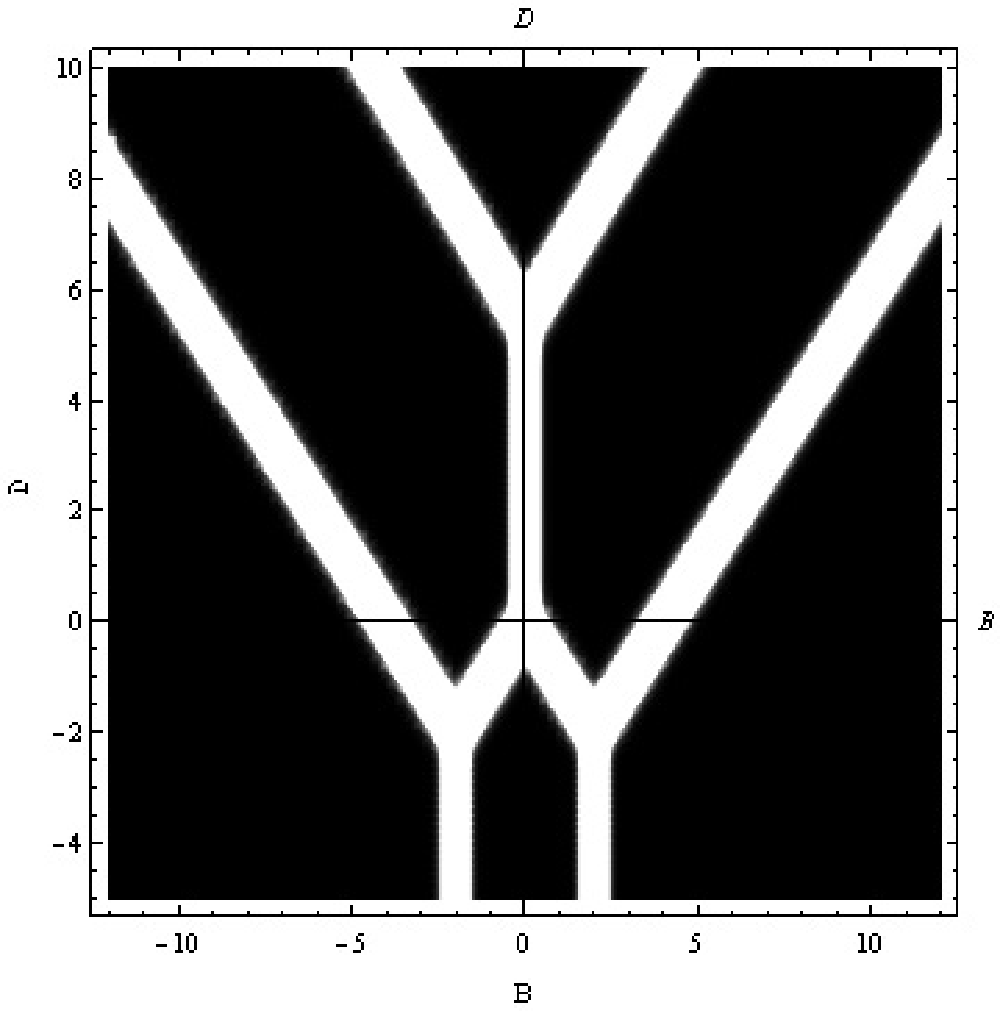}&\includegraphics[width=35mm]{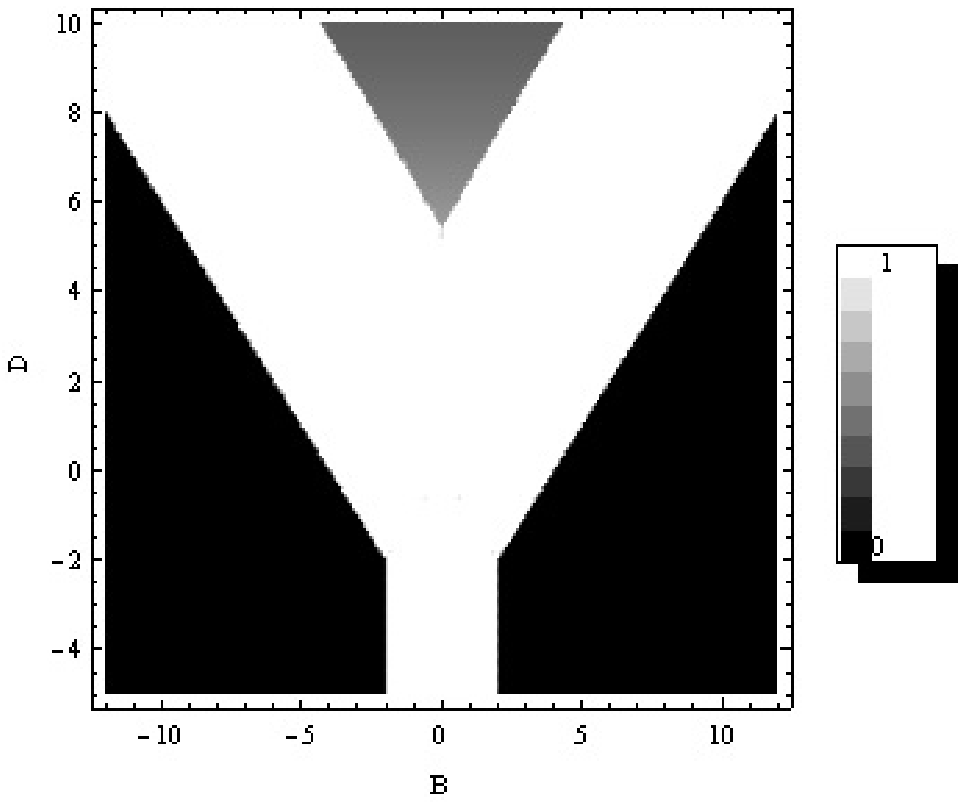}\\
\end{tabular}
\caption {\small {The density plots for (a) particle number $P$
(b) (quadrupole) particle susceptibility  (c) magnetic
susceptibility and (d) negativity versus $B$ and $D$ when $K=2$ in
antiferromagnetic case $J=1$.}}\label{BnonzeroKnonzero}
 \end{center}
\end{figurehere}
The distances along the magnetic field in figures~\ref{hbound03}
between various phases define the stable magnetic phases with
distinct spin gaps configurations, characterized by different spin
concentration and diverging susceptibilities along the boundaries.
For example, the negativity for white areas reaches the maximum
(saturated) value, while there are also different distinct areas
with partial (unsaturated) entanglement.
\begin{figurehere}
\begin{center}
\begin{tabular}{cc}
{\small (a)}&{\small (b)}\\
\includegraphics[width=35mm]{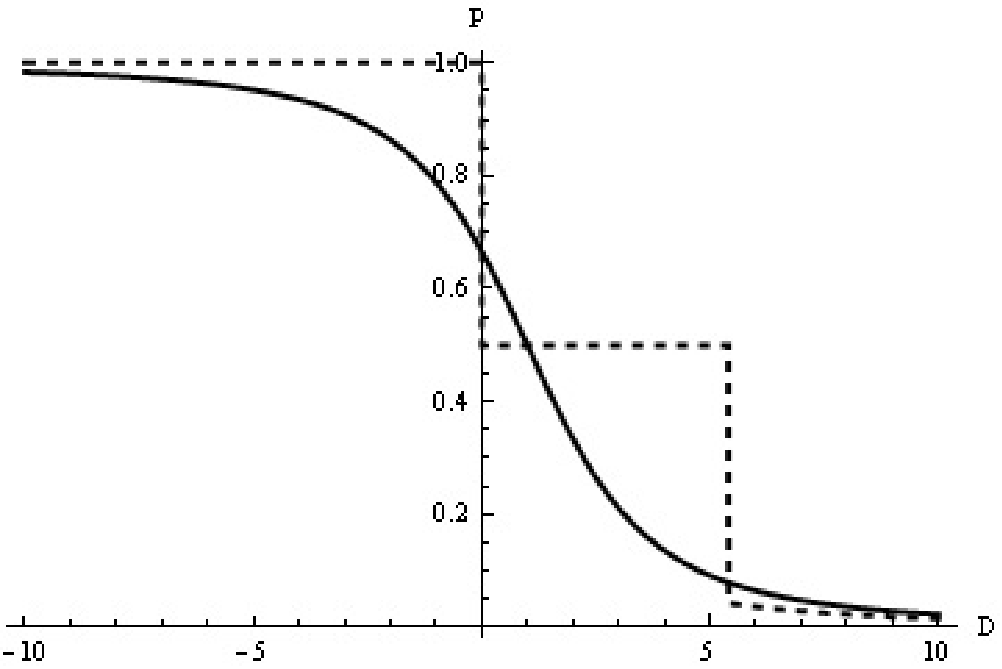}&\includegraphics[width=35mm]{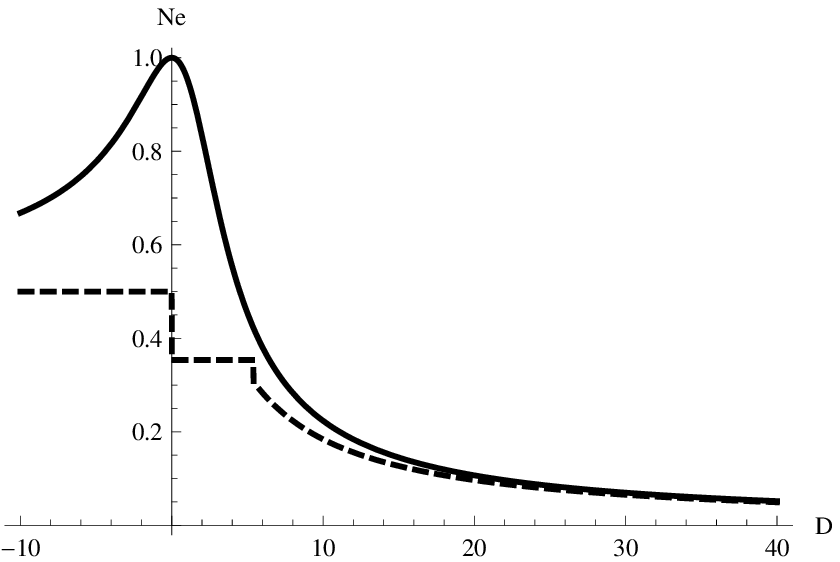}\\
\end{tabular}
\caption {\small {The density variation of a) particle number $P$
and b) negativity $Ne$ versus $D$ for $K=0$ (solid) and $K=2$
(dashed) for antiferromagnetic ($J=1$) in zero field $B=0$.}}
 \label{fig new}
 \end{center}
\end{figurehere}
These density plots can be used to determine of QCPs and the
boundaries for various QPTs. This result for finite size clusters
can have important consequences in the physics of quantum phase
transitions~\cite{Sach}, where so far the usual method to detect a
phase transition is to look at the scaling in the thermodynamic
systems. The competition among the different phases can lead to
complex behavior with the two triple points. The density of
negativity is an efficient indicator of QPTs. In
figure~\ref{BnonzeroKnonzero}(a) we find the new spin phase
boundaries from black to grey region with jump $\frac{1}{2}$ and
from grey region into the two white ones with the same jump. The
white middle line $B=0$ in figure~\ref{BnonzeroKnonzero}(c)
corresponds to classical effect at $J<0$ case (without change in
entanglement). On the other hand, the continuous lines seen in the
same plot at $J>0$ correspond to quantum phase transitions
(observable also in negativity).
\subsection{Effect of quadruple term}~\label{quadruple}
Here we display effect of nonlinear interactions between the
spins. The variation $P$ versus $D$ is shown in figure~\ref{fig
new}a for two quadruple interactions $K>0$ for antiferromagnetic
case with $J=1$. At $K=2$, an opposite spin pairing gap is opened
at $P=1/2$. Such a density profile, showing finite leap near
$P=1/2$, resembles the MH plateau behavior for the number of
particles versus chemical potential in the Hubbard clusters. This
is indicative of a possible opposite spin pairing
instability~\cite{PRB}. Therefore, the cluster behaves at large
$K$ as a MH like insulator in contrast to the spin liquid like
behavior with the zero gap, shown at $K=0$ in figure~\ref{fig
new}b. As it is seen from the density plot the magnetic field and
quadrupole interaction ($K$) makes the phase structure in
antiferromagnetic case more richer. Our analysis shows that the
negativity in $D-K$ space for ferromagnetic coupling is always
less than for antiferromagnetic case.
 \begin{figurehere}
           \begin{center}
             \psfig{figure=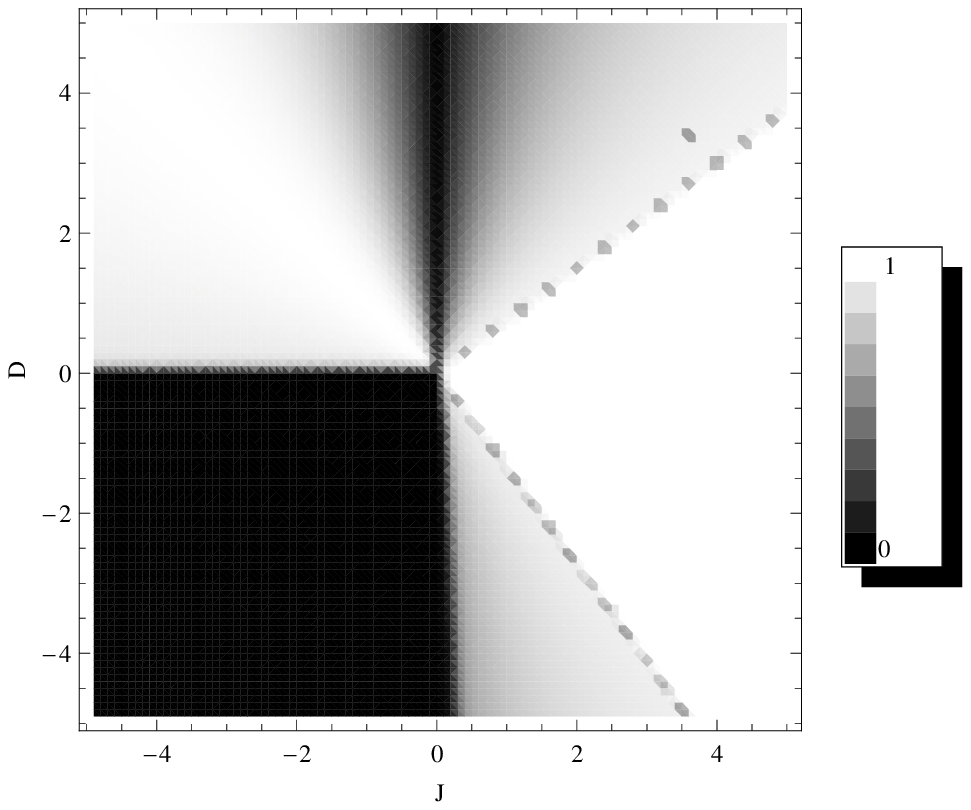,width=45mm}
           \end{center}
           \caption{\small{Density plot of negativity via $J$ and $D$ at $K=B=0$ in three-sites cluster.}}
           \label{3spinJD}\vspace{0.01cm}
         \end{figurehere}
\subsection{Non-bipartite clusters}~\label{NonBipartite}
Finally, we display in this session the results of negativity
versus $D$ and $J$ for frustrated three-site clusters in
figure~\ref{3spinJD} at rather low $T=0.01$. This picture for
ferromagnetic case resembles corresponding figure~\ref{hbound02}.
However, there is an apparent difference in behavior for the
region $J>0$, where there are two extra continuous boarder lines.
\bigskip
\section{Conclusion}~\label{conclusion}
In this paper we adopted the concept of entanglement to analyze
behavior of the small-size spin-1 Heisenberg clusters. We used a
negativity as a computable measure of entanglement to perform
extensive calculations of the negativity and response functions.
The critical fields and intrinsic parameters beyond which
entanglement disappears are calculated. We found regions where the
quantum entanglement can be increased more rapidly by increasing
both, $D$ and $K$. The negativity can determine the boarders
between ordered phases with excess correlations, above the
classical ones. The observed plateaus and peaks in spin and
particle behavior and susceptibilities can be considered as a
possible universal method for the simultaneous detection of
quantum and classical phase transitions. The (density) plots are
convenient (topographic map) tool for observation of the quantum
phases and quantum transitions. The states with vanishing
classical correlations but existing quantum correlations in
entanglement open up the new opportunities of phase transitions
that are detectable only through correlation in behavior of
entanglement and thermodynamic properties. Our studies of QCPs in
small size spin clusters appear to be generic to large
thermodynamic systems. The exact diagonalization is completely
unbiased for the study of QPTs and QCPs in strongly correlated
spin and electron systems~\cite{akocha}. Although the exact
studies have limitations (since the computations grow
exponentially with cluster size), we do not find a minimal
critical length in clusters below which a quantum critical
behavior disappears. The spin-1 boson Hubbard like model at
certain conditions can be mapped onto the spin-1 Heisenberg model.
Then these studies can also be
useful for the analysis of spontaneous phase separation and the
transition from Mott insulator to quantum superfluid in spin-1 Bose–Hubbard models on optical lattices.

\ack
We thank H Babujian for useful discussions. This work was
supported by ANSEF 2497-PS, 1981-PS and ECSP-09-08
and NFSAT research grants. LA thanks S Wimberger for
helpful discussion and acknowledges financial support from
the EMMI through a grant within the Institute of Theoretical
Physics of the University of Heidelberg, the Excellence
Initiative of the DFG through the Heidelberg Graduate School
of Fundamental Physics (grant number GSC 129/1). ANK thanks Edward Rezayi, Jose Rodriguez and Oscar Bernal for
helpful discussions.
% ----------------------------------------------------------------

\end{document}